\documentclass[twocolumn,pra,showpacs,superscriptaddress]{revtex4-1}
\usepackage{graphicx}
\usepackage{subfigure}
\usepackage{amsmath}
\usepackage{amsfonts,amssymb}
\usepackage{bm}
\usepackage{float}
\usepackage{color}
\usepackage{sidecap}
\allowdisplaybreaks
\usepackage{soul}
\setstcolor{red}
\usepackage{ulem}

\usepackage{hyperref}
\hypersetup{pdfpagemode=FullScreen,colorlinks=true,breaklinks,urlcolor=blue,linkcolor=blue,citecolor=blue}

\begin{document}

\title{Transport through a Non-Hermitian Aharonov-Bohm ring with physical gain and loss}

\author{Qi-Bo Zeng}
\email{zqb15@mails.tsinghua.edu.cn}
\affiliation{Department of Physics and State Key Laboratory of Low-Dimensional Quantum Physics, Tsinghua University, Beijing 100084, China}

\author{Shu Chen}
\affiliation{Beijing National Laboratory for Condensed Matter Physics, Institute of Physics,
Chinese Academy of Sciences, Beijing 100190, China}
\affiliation{Collaborative Innovation Center of Quantum Matter, Beijing 100084, China}

\author{Rong L\"u}
\email{rlu@tsinghua.edu.cn}
\affiliation{Department of Physics and State Key Laboratory of Low-Dimensional Quantum Physics, Tsinghua University, Beijing 100084, China}
\affiliation{Collaborative Innovation Center of Quantum Matter, Beijing 100084, China}

\begin{abstract}
We investigate a non-Hermitian Aharonov-Bohm (AB) ring system with a quantum dot (QD) embedded in each of its two arms. The energy levels of the QDs are complex in order to take into account the physical gain or loss of the ring system during its interacting processes with the environment. When there is magnetic flux threading through the ring, by allocating the flux phase factor into the tunneling amplitudes between the QDs and the leads in different ways, the Hamiltonian of the system can be written into different formalisms. We calculate the transmission through the ring by using these different non-Hermitian Hamiltonians and prove that it is not dependent on the way we treat the phase factor, as in the Hermitian case. In addition, with appropriate parameters, the asymmetric Fano profile will show up in the conductance spectrum just by tuning the physical gain and loss of the system. The Fano effect originates from the interferences of electrons traversing different channels which are broadened or narrowed down due to the interaction between the QDs and the environment. The proof we provide and the transport properties revealed in this paper demonstrate the influences of the environment on an otherwise isolated system and pave the way for the further studies on non-Hermitian AB ring systems.
\end{abstract}


\maketitle
\date{today}

\section{Introduction}

After nearly a century's development, the theory of quantum mechanics has become an indispensable component of modern science with so many highly accurate experimental verifications of its theoretical predictions. It is well known that in quantum mechanics, in order to guarantee the conservation of probability and keep the energy eigenvalues real, the Hamiltonian of a quantum system should be Hermitian \cite{Shankar}. To study the properties of a quantum system, we have to take into account of the interaction between the system and the environment. Normally the system we treat is localized in space and the environment can be considered to be a measuring device. However, there is always a natural environment which is independent of any observer and exists at all times. A quantum system thus should be treated as an open system by including the environment it is embedded \cite{IRotter}. Under these circumstances, quantum mechanics with non-Hermitian Hamiltonians can provide an effective way to include the influences of the environment on the system we discuss. Actually, many non-Hermitian Hamiltonians have been widely used to treat various problems in early days, such as free-electron lasers, transverse mode propagation in optical resonators and so on \cite{GDattoli, ASiegman, HCBaker, Moiseyev, Hatano1, Hatano2, Hatano3}. Non-Hermitian quantum mechanics attracted much more attention when people found that a large class of non-Hermitian Hamiltonians can exhibit all real eigenvalues when these systems are $\mathcal{PT}$-symmetric \cite{CMBender1, CMBender2}. These $\mathcal{PT}$-symmetric non-Hermitian systems are studied in many fields and have been experimentally realized in different physical systems in recent years \cite{Goldsheid, Heinrichs, Molinari, Regensburger, Guo, DKip, Chang, Peng, Fleury, Zhu, Schindler}.

The growing interest in non-Hermitian systems has motivated various discussions and extensions of the Hermitian Hamiltonians studied before, and have brought many more and deeper understandings about the quantum systems. For example, the topological properties of non-Hermitian systems are investigated in \cite{Liang, Esaki, Zeuner}. A non-Hermitian tight-binding network engineering method is proposed in \cite{Longhi1} and it is shown that effective complex non-Hermitian hopping rates can be realized with only complex onsite energies in the network. In \cite{Longhi2}, the author studied the spectral and dynamical properties of a quantum particle constrained on non-Hermitian quantum rings and found that very different behavior of particle motion showed up in the non-Hermitian case. The scattering propagation and transport problems in non-Hermitian systems have also been investigated and many interesting effects have been found \cite{Mostafa1, BZhu, Chong, Mostafa2, Ambichl, Ramezani, Jing, Garmon}.

Recently, the transmission through non-Hermitian scattering centers have been explored \citep{Li, Jin}. Interesting transport properties have been revealed in these non-Hermitian scatering systems. However, the authors did not check that whether it is still true that when the Aharonov-Bohm (AB) system is non-Hermitian, the transmission is still independent of the way we allocate the flux phase factor, which originates from the magnetic field threading through the AB ring, to the tunneling amplitudes between the AB ring and the leads, just like in the Hermitian systems. In addition, these papers are mainly focus on the $\mathcal{PT}$-symmetric cases, while in fact we can allocate the flux phase factor in different ways so that we can study the $\mathcal{PT}$ symmetric and asymmetric cases at the same time. Thus we should extend these systems to generalized non-Hermitian situations and check that if these different Hamiltonians with different flux factor allocation lead to the same result. 

\begin{figure}[!ht]
\centering
\includegraphics[width=3in]{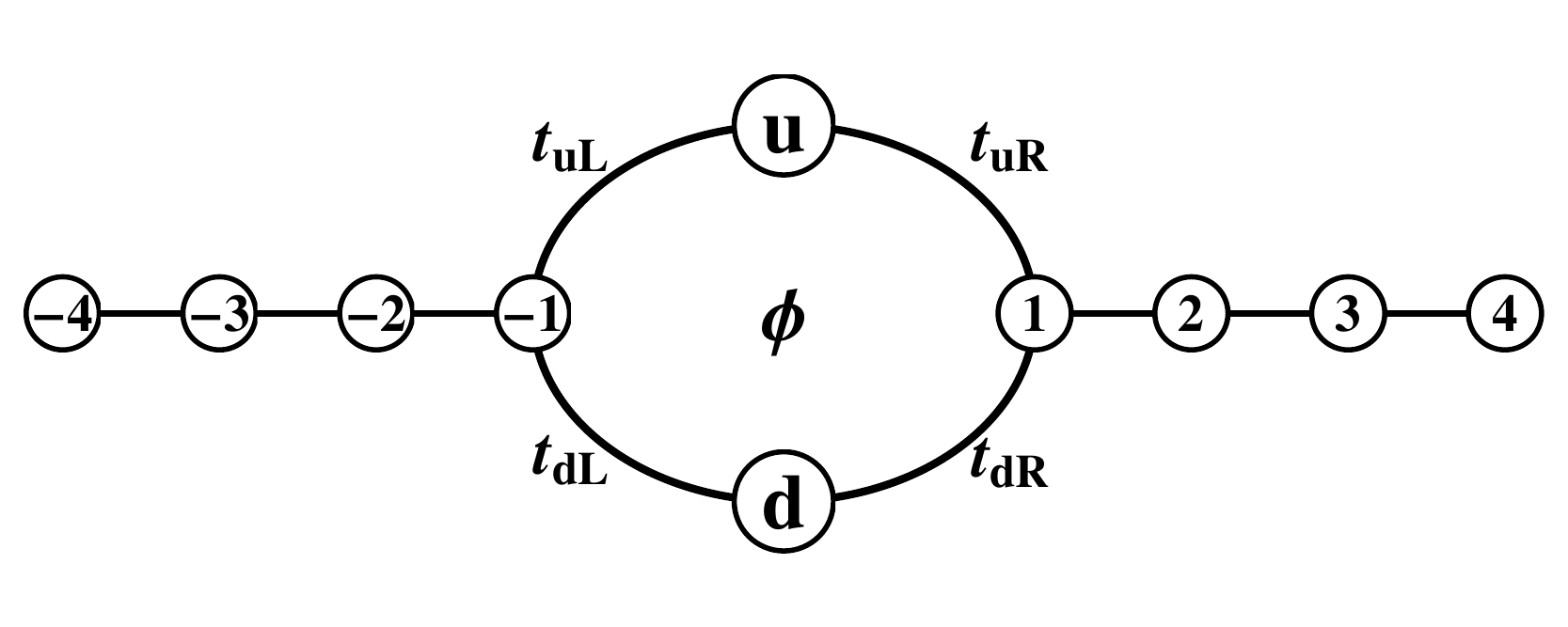}
\caption{Schematic setup of the Aharonov-Bohm ring system discussed in this paper. $u$ and $d$ are the two QDs embedded in the arms of the ring. The leads coupled to the ring are represented by two one-dimensional tight binding chains. $\phi$ is the magnetic flux threading through the ring.}
\label{fig1}
\end{figure}

In this paper, we study the transport properties through a non-Hermitian AB ring system as shown in Fig. \ref{fig1}. There are two quantum dots (QDs) embedded in the two arms of the ring and the ring is attached to two metallic leads which are represented by two one-dimensional chains. The energy levels of these two QDs can be complex in order to take into account of the physical gain or loss during the interacting processes between the ring and the environment. By allocating the flux phase factor induced by the magnetic flux threading through the ring into the tunneling amplitudes between the QDs and the leads in different ways, the Hamiltonian of the system would have different formalisms. We calculate the transmissions of the AB ring by using these Hamiltonians of the same system and find that they are equal to each other. The transmission is not dependent on the way we distribute the phase factor, which is the same as in the Hermitian case. This proof paves the way for further studies of the non-Hermitian AB ring systems. Besides, by checking the conductance spectrum, we find that the asymmetric Fano profile would show up by just tuning the physical gain and loss of the system. Due to the interaction between the QDs and the environment, the two channels through the AB ring will be broadened or narrowed down. Electrons traversing these channels with different widths will interfere and result in Fano effect in the conductance spectrum. This non-Hermitian system provides us with a simple model to check the influences of the environment on an otherwise isolated system.

The rest of the paper is organized as follows. In Sec. \ref{sec2}, we introduce the model Hamiltonians of the system. In Sec. \ref{sec3}, we calculate the transmissions through the AB ring according to the different Hamiltonians we introduced and compare these results. We also investigate the Fano profile in the conductance spectra of the system in this section. The last section (Sec. \ref{sec4}) is dedicated to a brief summary.

\section{Model and Hamiltonian}\label{sec2}
Consider an Aharonov-Bohm ring with a impurity site or quantum dot at each arm of the ring, see Fig. \ref{fig1}. We can add some imaginary potentials to the two QDs in the ring to represent the physical gain or loss during the interacting processes between the ring and the environment. The AB ring is also attached to two metallic leads which are represented by two one-dimensional chains. The Hamiltonian of such a system is
\begin{equation}\label{}
  \mathcal{H} = \mathcal{H}_{DQD} + H_{Leads} + H_T,
\end{equation}
where each part of $\mathcal{H}$ is described as follows
\begin{equation}\label{}
  \begin{split}
     \mathcal{H}_{DQD}=& E_u f_u^\dagger f_u + E_d f_d^\dagger f_d, \\
     \mathcal{H}_{Leads}=& -t_0 \sum_{j} (c_j^\dagger c_{j+1} + H.c.), \\
     \mathcal{H}_T= - & \sum_{n=u,d} (t_{nL} c_{-1}^\dagger f_n + t_{nR} c_1^\dagger f_n + H.c. ).
  \end{split}
\end{equation}
Here, $f_u^\dagger$ ($f_u$) and $f_d^\dagger$ ($f_d$) are the creation (annihilation) operators for the quantum dots implemented in the two arms of the ring with one single energy level $E_u$ and $E_d $ respectively. When $E_u$ and $E_d$ are both real, the Hamiltonian is Hermitian while if one or two of them are complex, the Hamiltonian becomes non-Hermitian. $c_j^\dagger$ ($c_j$) is the creation (annihilation) operator at site $j$ with $t_0$ being the hopping amplitude between the nearest sites in the chain. $t_{nL(R)}$ is the tunneling amplitude between the QDs and the lead $L$ ($R$). The tunneling amplitudes can be complex due to the allocation of the phase factor which originates from the magnetic filed threading through the ring. Due to gauge transformation, these phase factors can be allocated differently and leads to different Hamiltonians. In the Hermitian case, the physical variables will not be influenced by these differences, however, as we will show later, this physical picture also applies to this non-Hermitian system. We will mainly consider two kinds of Hamiltonians. One is symmetric with the phase factor distributed averagely to the four tunneling amplitudes
\begin{equation}\label{symcase}
 \begin{split}
  &t_{uL}= t e^{\frac{i\phi}{4}}, \hspace{0.5cm} t_{dL} = t e^{\frac{-i\phi}{4}},\\
  &t_{uR}= t e^{\frac{-i\phi}{4}}, \hspace{0.5cm} t_{dR} = t e^{\frac{i\phi}{4}}.
 \end{split}
\end{equation}
However, the Hamiltonian can also be written in an asymmetric form if the coupling strengths are chosen as
\begin{equation}\label{asymcase}
  t_{uL}=t e^{i\phi}, \hspace{0.5cm} t_{uR}=t_{dL}=t_{dR}=t.
\end{equation}
We will calculate the transmissions through the Aharonov-Bohm ring with these two different Hamiltonians and will compare the results with the Hermitian system in the following.

\section{Results and discussions}\label{sec3}
Now we calculate the transmission rate through the non-Hermitian AB ring with imaginary potentials. Suppose that the wave function of the system can be written as the linear combination of atomic orbitals, $| \Psi_k \rangle = \sum_n a_{nk} |n\rangle + \sum_{j} a_{jk} |j \rangle$, where $a_{nk}$ and $a_{jk}$ are the probability amplitudes to find the electrons with momentum $k$ at the QD site $n=u,d$ in the ring arms or at site $j$ in the leads, respectively. Assuming that there is an incoming electron from the left lead, and it is described by a plane wave which will be reflected and transmitted at the AB ring. Then we have
\begin{equation}\label{}
  \begin{split}
     a_{jkL} &= e^{ik\cdot j} + r_{LL} e^{-ik\cdot j}, \hspace{0.5cm} if \hspace{0.2cm} j<0 \\
     a_{jkL} &= \tau_{RL} e^{ik\cdot j}, \hspace{0.5cm} if \hspace{0.2cm} j>0,
  \end{split}
\end{equation}
where $r_{LL}$ and $\tau_{RL}$ are the reflection amplitude in the left lead and transmission amplitude from the left lead to the right lead, respectively. By substituting the wave function of the system into the Schr\"odinger equation $i \frac{\partial}{\partial t} |\Psi_k \rangle = H | \Psi_k \rangle$, we have
\begin{equation}\label{}
  \frac{\partial}{\partial t} |\Psi_k \rangle =  \dot{a}_{uk}|u\rangle + \dot{a}_{dk} |d\rangle + \sum_j \dot{a}_{jk} |j\rangle,
\end{equation}
and
\begin{equation}\label{}
\begin{split}
   & H |\Psi_k \rangle = E_u a_{uk} |u\rangle + E_d a_{dk} |d\rangle \\
     & -t_{uL} a_{uk} |-1\rangle - t_{uR} a_{uk} |1\rangle - t_{dL} a_{dk} |-1\rangle - t_{dR} a_{dk} |1\rangle \\
     & -t_{uL}^* a_{-1k} |u\rangle - t_{uR}^* a_{1k} |u\rangle - t_{dL}^* a_{-1k} |d\rangle - t_{dR}^* a_{1k} |d\rangle \\
     & -t_0 \sum_j a_{j-1,k} |j\rangle - t_0 \sum_{j} a_{j+1,k} |j\rangle.
\end{split}
\end{equation}
Let $a_{jk}(t) = a_{jk} e^{-i\omega t}$ with $\omega = -2t_0 \cos(k)$ being the energy dispersion of the one-dimensional chain. Then we have
\begin{equation}\label{}
  i \frac{\partial}{\partial t} |\Psi_k \rangle = \omega a_{uk} |u\rangle + \omega a_{dk} |d\rangle + \sum_j \omega a_{jk} |j\rangle.
\end{equation}
After substituting these into the Schr\"odinger equation, we can get the following equations
\begin{subequations}
\begin{equation}\label{a}
  -t_0 r_{LL} + t_{uL} a_{uk} + t_{dL} a_{dk} = t_0
\end{equation}
 \begin{equation}\label{b}
   -t_0 \tau_{RL} + t_{uR} a_{uk} + t_{dR} a_{dk} =0
 \end{equation}
 \begin{equation}\label{c}
    t_{uL}^* r_{LL} + t_{uR}^* \tau_{RL} + (\omega - E_u) e^{-ik} a_{uk} = -t_{uL}^* e^{-2ik}
 \end{equation}
 \begin{equation}\label{d}
    t_{dL}^* r_{LL} + t_{dR}^* \tau_{RL} + (\omega - E_d) e^{-ik} a_{dk} = -t_{dL}^* e^{-2ik}
 \end{equation}
\end{subequations}
From Eq. (\ref{a}) and (\ref{b}), we have
\begin{equation*}
  \begin{split}
     a_{uk} &= \frac{1}{A} t_0 [t_{dR} (1+r_{LL}) - t_{dL} \tau_{RL} ], \\
     a_{dk} &= \frac{1}{A} t_0 [-t_{uR} (1+ r_{LL}) + t_{uL} \tau_{RL} ],
  \end{split}
\end{equation*}
with $A$ defined as $A=t_{uL}t_{dR} - t_{dL} t_{uR}$. Substituting $a_{uk}$ and $a_{dk}$ into Eq. (\ref{c}) and (\ref{d}), we can get the transmission and reflection coefficient, which are shown as follows.
\begin{widetext}
\begin{equation}\label{}
\begin{split}
  &\tau_{RL} = \\
  &\frac{[ (\omega-E_u)e^{-ik} t_0 t_{dL}^* t_{dR} + (\omega-E_d)e^{-ik} t_0 t_{uL}^* t_{uR} ] (e^{-2ik}-1)}
  {A(t_{dL}^* t_{uR}^*-t_{uL}^* t_{dR}^*) - (\omega - E_u)e^{-ik} t_0 ( |t_{dL}|^2 + |t_{dR}|^2) - (\omega - E_d)e^{-ik} t_0 (|t_{uR}|^2 + |t_{uL}|^2) - (\omega - E_u)(\omega - E_d)e^{-2ik} t_0^2}
  \end{split}
\end{equation}
and
\begin{equation}\label{}
  r_{LL} = \frac{-A t_{uL}^* e^{-2ik} - (\omega - E_u)e^{-ik} t_0 t_{dR} - [ At_{uR}^* - (\omega - E_u) e^{-ik} t_0 t_{dL}] \tau_{RL}}
  {At_{uL}^* + (\omega - E_u) e^{-ik} t_0 t_{dR}}.
\end{equation}
When the phase factor is averagely distributed to the four hopping amplitudes, as shown in Eq. (\ref{symcase}), $A=2it^2 \sin \frac{\phi}{2}$, and the transmission coefficient becomes
\begin{equation}\label{}
  \tau_1= - \frac{[ (\omega - E_u) e^{i \frac{\phi}{2}} + (\omega - E_d) e^{-i\frac{\phi}{2}} ] (e^{-2ik} - 1) \Gamma}
  {(\omega - E_u)(\omega - E_d) e^{-ik} + 2\Gamma(\omega - E_u) + 2 \Gamma (\omega - E_d) + 4 \Gamma^2 e^{ik} \sin^2 \frac{\phi}{2} }
\end{equation}
where $\Gamma = \frac{t^2}{t_0}$.
However, if the phase factor is distributed as in Eq. (\ref{asymcase}), then $A=t^2 (e^{i\phi} -1)$, and the transmission coefficient becomes
\begin{equation}\label{}
  \tau_2 = - \frac{[ (\omega - E_u) + (\omega - E_d) e^{-i \phi} ] (e^{-2ik} - 1) \Gamma}
  {(\omega - E_u)(\omega - E_d) e^{-ik} + 2\Gamma(\omega - E_u) + 2 \Gamma (\omega - E_d) + 4 \Gamma^2 e^{ik} \sin^2 \frac{\phi}{2} }.
\end{equation}
Apparently, $\tau_2=e^{-i\phi/2} \tau_1$, so the transmissions through the AB ring system, which is defined as $T=|\tau|^2$, will be the same for the different Hamiltonians. The transmission can be expressed as
  \begin{equation}\label{T}
    T = \frac{1}{|B|^2} 4 \Gamma^2 \sin^2 k [ (\omega - E_u)(\omega - E_u^*) + (\omega - E_u)(\omega - E_d^*)e^{i\phi} + (\omega - E_u^*)(\omega - E_d)E^{-i\phi} + (\omega - E_d)(\omega - E_d^*) ],
  \end{equation}
where $B=(\omega - E_u)(\omega - E_d) e^{-ik} + 2\Gamma(2\omega - E_u - E_d) + 4 \Gamma^2 e^{ik} \sin^2 (\phi/2)$ is the denominator of the transmission amplitude. So the transmission is not dependent on how we distribute the phase factors in the Hamiltonian even when the system is non-Hermitian.

Now, let's consider the non-Hermitian cases with and without $\mathcal{PT}$-symmetry. If $E_u= \epsilon + i\gamma$ and $E_d=\epsilon - i\gamma$, the Hamiltonian is $\mathcal{PT}$-symmetric when the phase factor is written in the form in Eq. (\ref{symcase}), the transmission of the system is
\begin{equation}\label{}
  T_1=\frac{1}{|B|^2} 4 \Gamma^2 \sin^2 k \{ 2[(\omega - \epsilon)^2 + \gamma^2] + 2\cos \phi (\omega - \epsilon)^2 + 4\gamma \sin \phi (\omega - \epsilon) - 2 \gamma^2 \cos \phi  \}.
\end{equation}
However, if the Hamiltonian is written in an form without $\mathcal{PT}$-symmetry, as in Eq. (\ref{asymcase}), the transmission of the system becomes
\begin{equation}\label{}
  T_2=\frac{1}{|B|^2} 4 \Gamma^2 \sin^2 k \{ 2[(\omega - \epsilon)^2 + \gamma^2] + 2\cos \phi (\omega - \epsilon)^2 + 4\gamma \sin \phi (\omega - \epsilon) - 2 \gamma^2 \cos \phi  \},
\end{equation}
with $B=[(\omega - \epsilon)^2 + \gamma^2] e^{-ik} + 4\Gamma(\omega - \epsilon) + 4\Gamma^2 e^{ik} \sin^2(\phi/2)$. Thus $T_1 = T_2$, the transmissions calculated by using $\mathcal{PT}$-symmetric and -asymmetric Hamiltonian are the same.
\end{widetext}

In fact, if only one of the two QDs is coupled by imaginary potential, namely only $E_u$ or $E_d$ is complex, the transmissions calculated by those different Hamiltonians would also become equal with each other, as shown in Eq. (\ref{T}), though the Hamiltonian can not be $\mathcal{PT}$-symmetric any more. So as long as there is imaginary potential added to the QDs in the arms of the AB ring, the transmission of the system we get will not depend on the form of Hamiltonian we write down.

Next let's investigate the conductance properties of the system. The conductance through the AB ring is defined as $G=\frac{2e^2}{h} T$. Here we mainly focus on the conductance at the Fermi energy ($k=\pi/2$ then $\omega=0$) and we have
\begin{equation}\label{}
  \tau(k=\frac{\pi}{2}) = -i \frac{2\Gamma (E_u e^{i \frac{\phi}{2}}+E_d e^{-i\frac{\phi}{2}})}{(E_u - 2i\Gamma)(E_d -2i\Gamma)+4\Gamma^2 \cos^2 \frac{\phi}{2}}.
\end{equation}
This is very similar to the expression in \cite{Agundez} except that the energy levels of the QDs are complex now. Due to the coupling to the leads, the level of the QD will be broadened and the width is represented by $2\Gamma$. Since $E_u$ and $E_d$ are complex, we can set $E_{u/d} = \epsilon_{u/d} + i \gamma_{u/d}$ with $\epsilon_{u/d}$ and $\gamma_{u/d}$ being real, then the width of the energy levels for the $u$ dot and the $d$ dot are $(2\Gamma-\gamma_u)$ and $(2\Gamma-\gamma_d)$, respectively. So if $\gamma_u \neq \gamma_d$, these two levels will have different widths, one is broader and the other is relatively narrow. Then there are two channels in this system, with appropriate parameters, the broader channel can be taken as a continuous background while the narrow one as a discrete resonant channel. Electrons traversing through these two different channels will interfere with each other and lead to the asymmetric Fano profile in the conductance spectra \cite{Fano, Mirosh}. We have supposed that the dots are symmetrically coupled to the leads, so the only way to differently change the widths of the two QD energy levels is by tuning $\gamma$, namely by tuning the physical gain or loss originating from the interaction between the QDs and the environment. The situation becomes more clear when $\gamma_u$ is positive while $\gamma_d$ is negative, since then one of the energy level will be broadened while the other will be narrowed down, thus the interferences of electrons traveling through these two channels will make the Fano profile more significant.

\begin{figure}[!ht]
\centering
\includegraphics[width=3in]{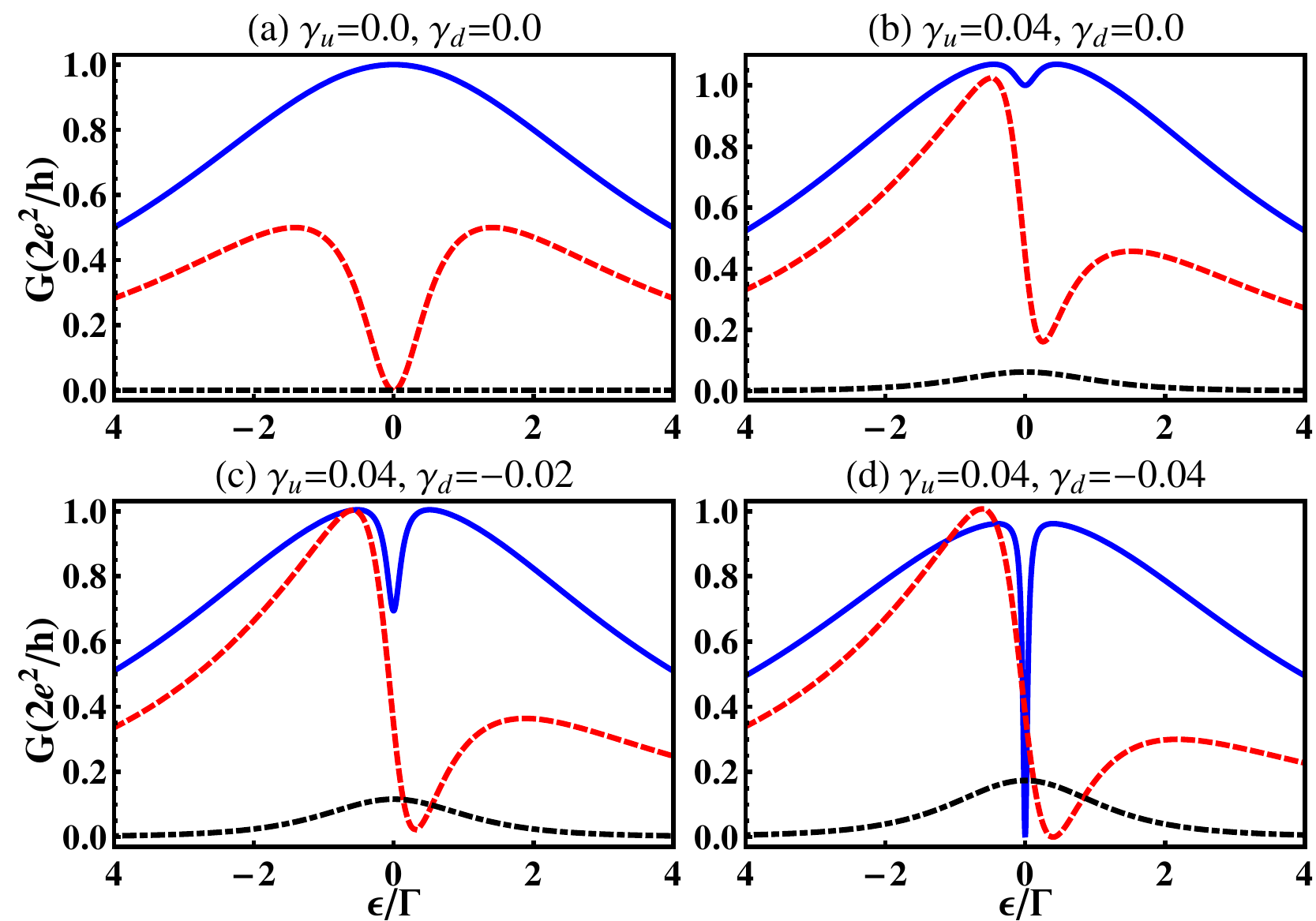}
\caption{(Color online) The conductance through the AB ring system with different physical gain and loss when $\phi=0.0$ (blue solid curve), $\phi=0.5\pi$ (red dashed curve) and $\phi=\pi$ (black dot-dashed curve). }
\label{fig2}
\end{figure}

In Fig. \ref{fig2}, we present the conductance spectra of the AB ring system with different physical gain and loss. We take $t_0=1$ as the energy unit and set $\Gamma=0.1t_0$ throughout this paper. When there is no physical gain or loss in the system, the conductance spectra are always symmetric (Fig. \ref{fig2}(a)). When $\gamma_u \neq \gamma_d$, as shown by the red dashed curve in Fig. \ref{fig2}(b), (c) and (d), the asymmetric Fano lineshape shows up when $\phi=\pi/2$. The Fano profile becomes more sharp when the difference between $\gamma_u$ and $\gamma_d$ gets larger. Besides, there is a dip when $\phi=0$, which is denoted by the blue solid line and the dip will reach to zero when $\gamma_u=\gamma_d$ (see Fig. \ref{fig2}(d)). When $\phi=\pi$, the conductance keeps the symmetric Lorentzian shape (or zero) in situations with (or without) physical gain and loss, as represented by the black dot-dashed line in the figure.

Actually, if we choose $E_u=E_d=\epsilon$ and $\gamma_u = \gamma_d=\gamma$, which corresponds to a situation with balanced physical gain and loss in these two QDs. Then the conductance of the system becomes
\begin{equation}\label{G}
  G=\frac{2e^2}{h}\cos^2 \frac{\phi}{2} \frac{(\epsilon - \gamma \tan \frac{\phi}{2})^2}{\epsilon^2 + \alpha^2},
\end{equation}
where $\alpha=\frac{\epsilon^2 - (4\Gamma^2 \sin^2 \frac{\phi}{2} - \gamma^2)}{4\Gamma}$. This is very similar to the standard formula for Fano resonance profile which is defined as \cite{Fano, Mirosh}
\begin{equation}\label{}
  \sigma = \frac{(\epsilon + q)^2}{\epsilon^2 +1},
\end{equation}
where $q$ is the asymmetric parameter. We can simply take $q=-\gamma \tan (\phi/2)$ in Eq. (\ref{G}). When $\phi=0$, $q=0$, the conductance profile is symmetric and there is a dip down to zero, as shown by the blue solid line in Fig. \ref{fig2}(d). When $\phi=\pi$, $q=-\infty$, the conductance shows the standard symmetric Lorentzian type, like the black dot-dashed curve shows. When $\phi=0.5\pi$, we have $q=-\gamma$,   and we have the asymmetric Fano profile and the dip shows up at $\epsilon=-q$, just as the red dashed line indicates. Though we do not normalize the conductance, all the characteristics revealed in the conductance spectrum of our system are consistent with the standard Fano profiles. The difference about this non-Hermitian system is that the broadened width of the QD energy level and thus the width of the channel are controlled by the physical gain and loss of the system. So the influences of the environment are directly reflected in the behavior of the conductance spectrum.

Another aspect needs to be noticed is that the maximum of the conductance will exceed $2e^2/h$ when the $\gamma_{u/d}$ becomes large, which denotes that the probability is not conserved in this system due to the non-Hermiticity.

\section{Summary}\label{sec4}
We investigate a non-Hermitian Aharonov-Bohm ring system in which the energy level of the two embedded quantum dots (QDs) in the two arms of the ring could be complex. The complex energy level represents the physical gain or loss during the interacting process between the AB ring and the environment. Due to the magnetic flux threading through the ring, the Hamiltonian of this model can be written in different forms by differently distributing the phase factor inducing by the magnetic field to the hopping amplitudes between the QDs and the leads. We calculate the transmission through the AB ring using these different Hamiltonians, including $\mathcal{PT}$-symmetric and -asymmetric cases, and find that it is not dependent on the way we distribute the phase factor in the Hamiltonian, which is the same as in the Hermitian case. In addition, by checking the conductance spectrum, we find that the asymmetric Fano profile can show up by just tuning the physical gain and loss of the system. The interaction between the QDs and the environment will broaden or narrow down the two channels through the ring and electrons traveling through different channels will interfere and result in Fano effect. So the influence of the environment is revealed in the transport properties of the system. This non-Hermitian Aharonov-Bohm ring system we discussed in this paper provides a simple model to check the influence of the environment on an otherwise isolated system and a demonstration of the basic principles of quantum mechanics. The proof we provide here, however, paves the way for further studies on non-Hermitian AB ring systems.

\section*{Acknowledgments}
This work has been supported by the NSFC under Grant No. 11274195 and the National Basic Research Program of China (973 Program) Grant No. 2011CB606405 and No. 2013CB922000. Shu Chen is supported by NSFC under Grants No. 11425419, No. 11374354 and No. 11174360, and the Strategic Priority Research Program (B) of the Chinese Academy of Sciences (No. XDB07020000).

\end{document}